%% file: prl.tex
\documentclass[aps,prl,twocolumn,showpacs,superscriptaddress]{revtex4}  

\usepackage{graphicx}  
\usepackage{dcolumn}   
\usepackage{bm}        
\usepackage{amssymb}   
\usepackage{epsfig}
\usepackage{psfrag}
\usepackage{subfigure}
\usepackage{xcolor}
\def\tev{\ifmmode {\mathrm{~Te\kern -0.1em V}}\else \textrm{~Te\kern -0.1em V}\fi}%
\def\gev{\ifmmode {\mathrm{~Ge\kern -0.1em V}}\else \textrm{~Ge\kern -0.1em V}\fi}%
\def\ifb{\mbox{ fb$^{-1}$}}
\newcommand{\Eslash}{\mbox{$E \kern-0.6em\slash$                }}
\newcommand{\met}{\mbox{\ensuremath{\Eslash_{\kern-0.3emT}\!}                        }}

\def\llbb{$ZH\to\ell^+ \ell^- b\bar{b}$}
\def\vvbb{$ZH\to\nu\bar{\nu} b\bar{b}$}
\def\lvbb{$WH\to\ell\nu b\bar{b}$}

\begin{document}

\hspace{5.2in} \mbox{FERMILAB-PUB-12-406-E}

\title{Combined search for the standard model Higgs boson\\ decaying to \boldmath ${b\bar{b}}$ \unboldmath using the D0 Run II data set}

\input author_list.tex       
\date{July 29, 2012}

\begin{abstract}
We present the results of the combination of searches for the standard
model Higgs boson produced in association with a $W$ or $Z$ boson and
decaying into $b\bar{b}$ using the data sample collected with the D0
detector in $p\bar{p}$ collisions at $\sqrt{s}=1.96\tev$ at the
Fermilab Tevatron Collider.  We derive 95\% C.L. upper limits on the
Higgs boson cross section relative to the standard model prediction in
the mass range $100\gev\leq M_H \leq 150\gev$, and we exclude Higgs
bosons with masses smaller than $102\gev$ at the 95\% C.L. In the mass
range $120\gev\leq M_H \leq 145\gev$, the data exhibit an excess above
the background prediction with a global significance of $1.5$ standard
deviations, consistent with the expectation in the presence of a
standard model Higgs boson.
\\
\\
DOI: 10.1103/PhysRevLett.109.121802
\end{abstract}

\maketitle

Despite its success as a predictive tool, the standard model (SM) of
particle physics~\cite{sm} remains incomplete without a means to
explain electroweak symmetry breaking. The simplest proposed
mechanism~\cite{ewsb} involves the introduction of a complex doublet
of scalar fields that generates the masses of elementary particles via
their mutual interactions.  After accounting for longitudinal
polarizations for the electroweak bosons, this mechanism also gives
rise to a single scalar boson, the SM Higgs boson, with an unpredicted
mass ($M_H$).  Direct searches for $e^{+} e^{-} \rightarrow
Z^{*}\rightarrow ZH$ at the CERN $e^+e^-$ Collider (LEP) yielded a
lower mass limit of $M_H >114.4\gev$~\cite{lepcombo} at 95\%
confidence level (C.L.).  Precision electroweak
measurements~\cite{lepewwg}, including the latest $W$ boson mass
measurements~\cite{CDFMW,DZMW} at the Fermilab Tevatron Collider,
result in an upper 95\% C.L. limit of $M_H < 152\gev$.  Direct
searches at LEP~\cite{lepcombo}, the Tevatron~\cite{tevcomb2012}, and
the CERN Large Hadron Collider (LHC)~\cite{atlascomb2012,cmscomb2012}
exclude at the 95\% C.L. most of the allowed mass range, except for
$116.6\gev < M_H <119.4\gev$ and $122.1\gev < M_H <127.0\gev$. In
addition, the ATLAS and CMS Collaborations have
published~\cite{atlascomb2012,cmscomb2012} excesses above background
expectations at a mass of $\approx 125\gev$ and have recently
published results~\cite{sm-lhc-ichep} confirming these excesses at the
level of 5 standard deviations (s.d.), driven by searches for $H \to
\gamma\gamma$ and $H\to Z Z^{(*)}\to \ell^+ \ell^- \ell^{\prime
  +}\ell^{\prime -}$, where $\ell$ and $\ell^{\prime}$ denote an
electron or muon.  These searches primarily exploit the gluon-gluon
fusion production mechanism for the Higgs boson, $gg \to H$, mediated
by a top-quark loop, while $H \to \gamma\gamma$ searches are also
sensitive to vector ($V=W,Z$) boson fusion, $q\bar{q^\prime}\to H
q\bar{q^\prime}$.  In the allowed mass range, the Tevatron experiments
are particularly sensitive to the SM Higgs boson produced in
association with a vector boson, $VH$, and the Higgs boson decaying
into $b\bar{b}$, the primary decay mode for a Higgs boson with $M_H <
135\gev$.  Searches at both hadron colliders have a high degree of
complementarity, with the main search channels at the LHC being
particularly sensitive to the Higgs boson mass and couplings to vector
bosons, while searches at the Tevatron provide information on the
Higgs boson coupling to $b$ quarks.

This Letter describes the combination of searches for $VH$, $H\to
b\bar{b}$ production at the D0 experiment using the sample of
$p\bar{p}$ collision data at $\sqrt{s}=1.96\tev$ collected during Run
II of the Fermilab Tevatron Collider.  These searches are focused on
leptonic $W$ and $Z$ boson decays that allow us to efficiently
suppress the large multijet background present at a hadron collider
and are restricted to the mass range $100\gev\leq M_H \leq
150\gev$. Therefore, the signal processes being targeted are
\lvbb~\cite{whlvbb}, \vvbb~\cite{zhvvbb}, and \llbb~\cite{zhllbb}.  A
similar combination of searches in the $H\to b\bar{b}$ decay mode has
recently been reported by the CDF Collaboration~\cite{cdfbb} and
previously by the ATLAS~\cite{atlasbb}, CMS~\cite{cmsbb}, and
LEP~\cite{lepcombo} Collaborations.

The D0 detector is described elsewhere~\cite{d0det}. Details on the
reconstruction and identification criteria for the physics objects
used in these searches [electrons, muons, jets, and missing transverse
energy (\met)] can be found
elsewhere~\cite{whlvbb,zhvvbb,zhllbb,whprd}.  Jets are identified as
consistent with the fragmentation of a $b$ quark ($b$-tagged) by a
multivariate algorithm~\cite{btagnim} combining information from the
impact parameter of displaced tracks and the topological properties of
secondary vertices reconstructed in the jet.

The main backgrounds affecting these searches originate from
$W/Z$+heavy-flavor jets (jets initiated by $b$ and $c$ quarks) and
from top-quark pair ($t\bar{t}$) production. Smaller contributions
arise from $W/Z$+light-flavor jets, single top-quark, diboson
($WW,WZ,ZZ$), and multijet production.  Multijet events contribute to
the selected samples via the misidentification of a jet or a photon as
an electron, the presence of a non-prompt lepton from a semileptonic
$b$- or $c$-hadron decay (\lvbb\ and \llbb\ analyses), or jet
energy mismeasurements resulting in apparent large \met (\vvbb\
analysis). In all instances, the normalization and kinematic
distributions of multijet events are estimated via data-driven
methods.  The remaining backgrounds, as well as the signal, are
estimated with Monte Carlo simulation.  Samples of $W/Z$+jets and
$t\bar{t}$ events are generated by using the {\sc alpgen}~\cite{alpgen}
tree-level matrix element generator, while samples of single top-quark
and diboson events are generated by using the {\sc
SingleTop}~\cite{singletop} and {\sc pythia}~\cite{pythia}
leading-order (LO) generators, respectively.  These samples are
normalized to next-to-next-to-LO (NNLO)~\cite{vjetsxs}, approximate
NNLO~\cite{topxs,stopxs}, and next-to-LO~\cite{mcfm} theoretical
cross sections.  Samples of $WH$ and $ZH$ signal events are generated
by using the {\sc pythia} generator for a range of masses, $100\gev\leq
M_H \leq 150\gev$, in steps of $5\gev$ and are normalized to the most
recent theoretical predictions~\cite{vhxs,hdecay,pro4f}. All Monte Carlo
samples are generated by using the {\sc CTEQ6L1} PDF set~\cite{cteq6} and
processed through {\sc pythia} to model parton showering and
fragmentation.  Signal and backgrounds samples are processed by a {\sc
geant3}-based~\cite{geant} simulation of the D0 detector and
reconstructed by using the same algorithms applied to the collider data.
Simulated events are corrected so that the object identification
efficiencies, energy scales, and energy resolutions match those
determined in data control samples.  More details on the simulation
and normalization of the signal and background samples can be found
elsewhere~\cite{whlvbb,zhvvbb,zhllbb}.

In the case of the $ZH \to \nu\bar{\nu} b\bar{b}$ analysis, the data
were collected by using triggers requiring jets plus \met and
correspond to an integrated luminosity of $9.5\ifb$~\cite{d0lumi}.
The \llbb\ and \lvbb\ analyses use a logical OR of triggers dominated
by single lepton, dilepton, lepton-plus-jets, and jet-plus-\met
triggers, resulting in an integrated luminosity of $9.7\ifb$. The
analyses select non-overlapping subsets of data via different
requirements on lepton multiplicity: (i) exactly two opposite-charge
leptons (\llbb), (ii) exactly one charged lepton and large \met
(\lvbb), and (iii) exactly zero charged leptons and large \met
(\vvbb). A significant fraction of signal events selected by the
\vvbb\ analysis originate from $WH$ production, where the charged
lepton is not reconstructed. In addition, events are required to have
two or three reconstructed jets, with the exception of the
\vvbb\ analysis, which is restricted to events with exactly two
jets. The signal-to-background ratio is significantly enhanced by
requiring one or two $b$-tagged jets in an event. The sensitivity of
the searches is maximized by categorizing events into different
analysis subchannels depending on the flavor and quality of the
charged leptons, jet multiplicity, $b$-tagged jet multiplicity, and
$b$-tagged jet quality.  The primary discriminating variable between
the $VH$ signal and the backgrounds is the dijet invariant mass, for
which the signal shows a distinct resonant structure; however, by
combining this variable with several other kinematic variables via a
multivariate approach, the sensitivity of the searches is improved by
approximately 25\%.  Therefore, the final observable for each of the
subchannels in the different searches is a one-dimensional
multivariate discriminant optimized for each hypothesized $M_H$ value.

We interpret the result of the searches via the ${\rm CL}_{\rm s}$
method~\cite{CLs-1,CLs-2}, which employs a log-likelihood ratio ${\rm
  LLR}=-2\ln (L_{\rm s+b}/L_{\rm b})$ as a test statistic, where
$L_{\rm s+b}$ ($L_{\rm b}$) is a Poisson likelihood to observe the
data under the signal-plus-background (background-only) hypothesis.
Separate channels are combined by summing LLR values over all bins,
thus maintaining the individual channel sensitivities.  The per-bin
signal and background predictions are parameterized in terms of
nuisance parameters that describe the effect of systematic
uncertainties.  The impact of systematic uncertainties on the search
sensitivity is reduced by maximizing both likelihood functions
$L_{\rm s+b}$ and $L_{\rm b}$, with respect to these nuisance
parameters, subject to Gaussian constraints of their prior
values. ${\rm CL}_{\rm s}$ is defined as the ratio of the confidence
levels for the signal-plus-background (${\rm CL}_{\rm s+b}$) and
background-only (${\rm CL}_{\rm b}$) hypotheses, which are each
evaluated by integrating the corresponding LLR distributions populated
by simulating outcomes via Poisson statistics.  Systematic
uncertainties are incorporated via Gaussian fluctuations on the
expected number of signal and background events per bin, taking into
account correlations across processes and
channels~\cite{correlations}.  Signal cross sections
resulting in ${\rm CL}_{\rm s}<0.05$ are excluded at the 95\%~C.L.

The systematic uncertainties differ between analyses, but we summarize
here the largest contributions.  We account for the impact of these
uncertainties both on the integrated signal and background yields and
on the shapes of the final discriminants where relevant.  The
\vvbb\ and \lvbb\ analyses carry a correlated uncertainty on the
integrated luminosity of 6.1\%~\cite{d0lumi}.  The \llbb\ analysis
normalizes the predictions using the peak from $Z\rightarrow \ell^+
\ell^-$ decays from data and the corresponding NNLO cross
section~\cite{vjetsxs}.  The $b$-tagging efficiency has an uncertainty
of $\approx$1\%--15\%, depending on the sample and $b$-tagging criteria.
The uncertainty due to acceptance and energy measurement of jets is
typically around 7\%.  Uncertainties due to acceptance and energy
measurement of leptons range from 1\% to 9\%, depending on the final
state.  A significant source of uncertainty comes from the $V+$jets
background cross sections, which have uncertainties of 4\%--10\% for
light flavor jets and $\approx 22\%$ for heavy flavor jets.  These
account for both the uncertainty on the theoretical cross section
calculations and the uncertainties on the higher-order correction
factors.  The uncertainty on the expected multijet background is
dominated by the statistics of the data sample from which it is
estimated, and is considered separately from the other cross section
uncertainties.  All analyses take into account the uncertainties on
the theoretical production cross sections for the different signal
processes due to PDF and scale choice.  In addition, analyses
incorporate differential uncertainties on the dominant backgrounds to
allow for potential variations of the final discriminants due to
generator and background modeling uncertainties.  The total impact of
systematic uncertainties on the combined sensitivity is $\approx20\%$.

\begin{figure}[h!]
\centering {\includegraphics[width=\columnwidth]{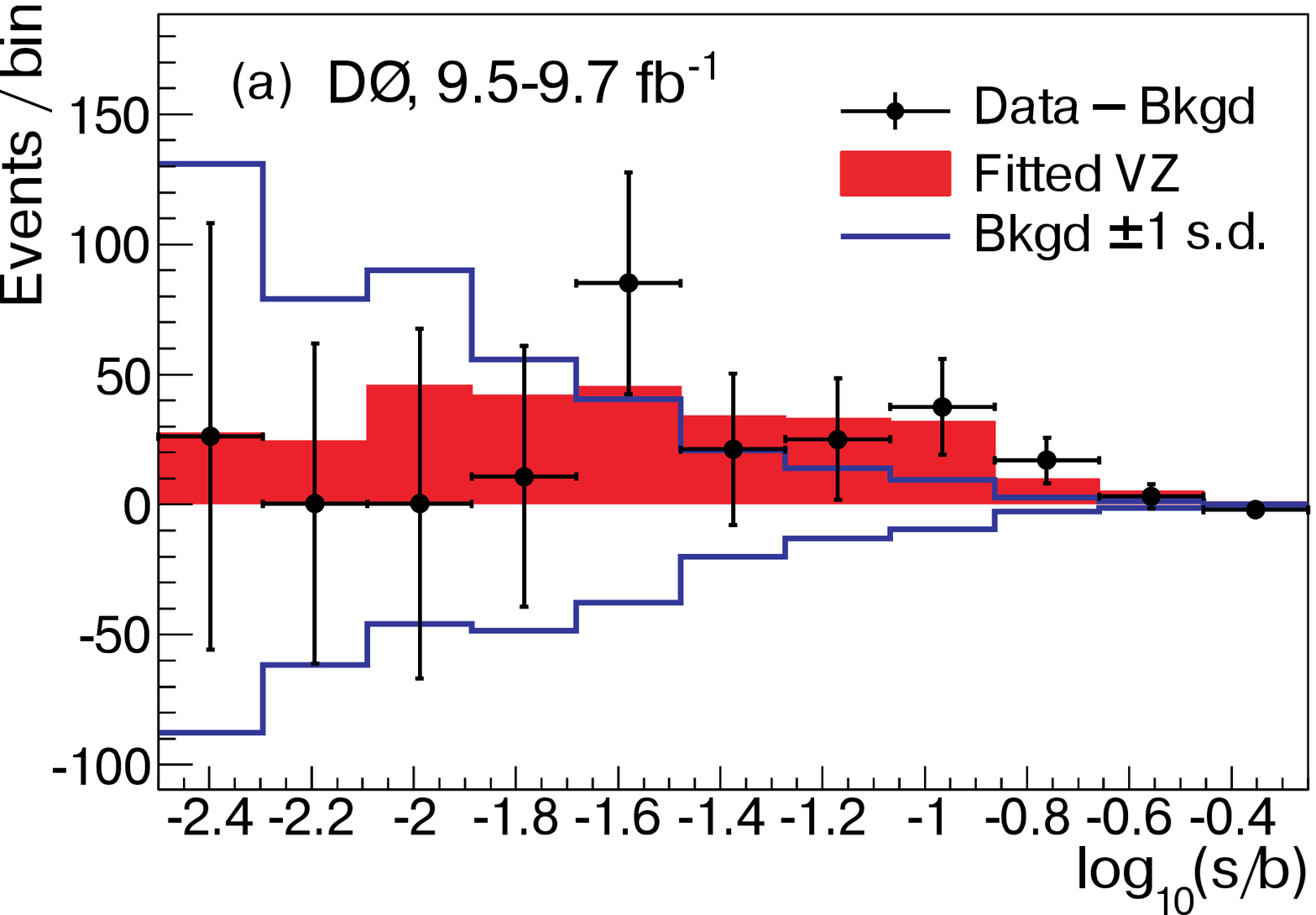}}
           {\includegraphics[width=\columnwidth]{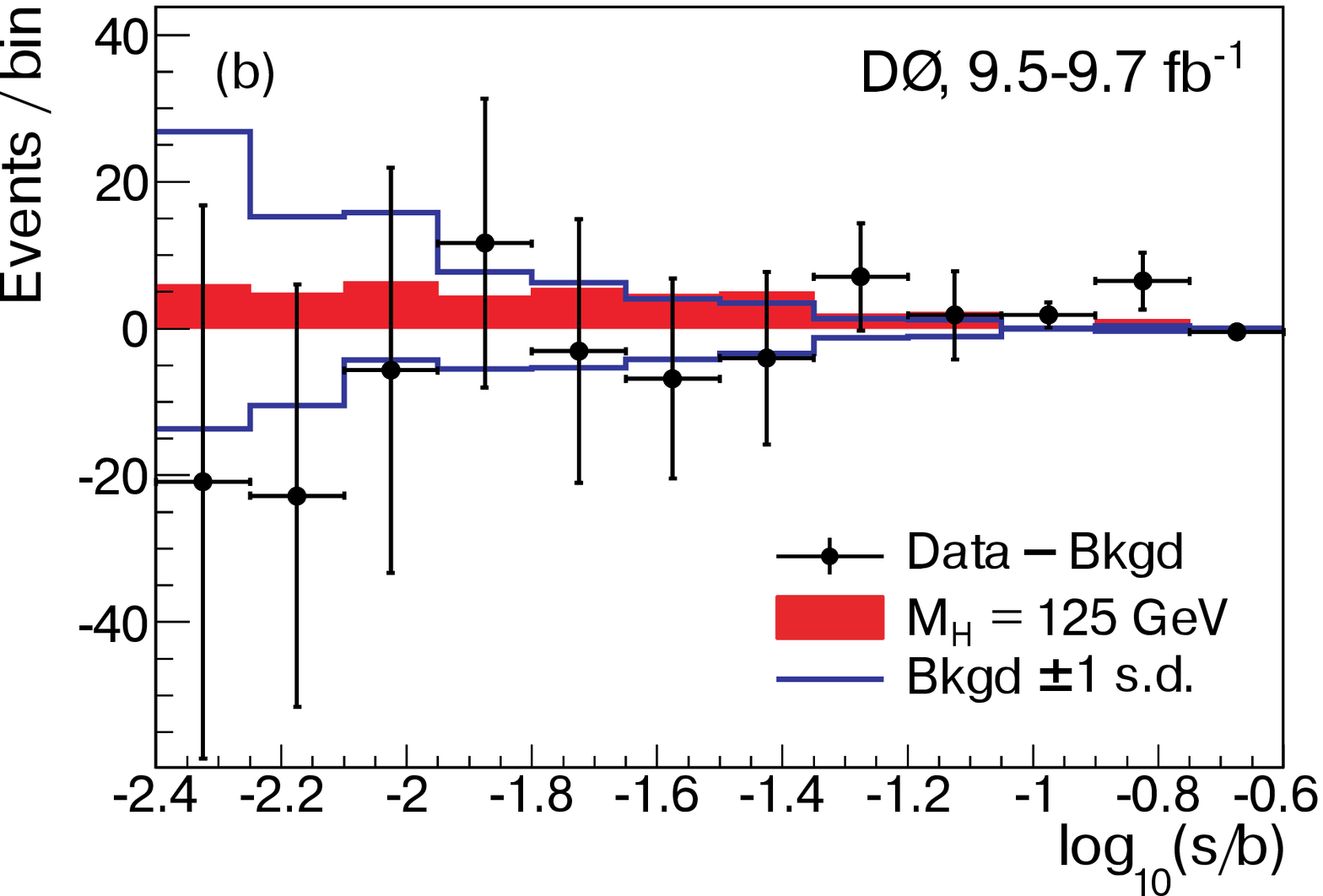}}
\caption{\label{fig:compare_subtracted} (color
  online). Background-subtracted data distributions of
  $\log_{10}(s/b)$ in (a) the $VZ$ analysis after a fit of the $VZ$
  and background contributions to the data and (b) the $VH$, $H\to
  b\bar{b}$ search for $M_H = 125\gev$ after a fit of the backgrounds
  to the data.  The background-subtracted data are shown as points and
  the signal is shown as the red histogram in each plot.  The blue
  lines indicate the posterior uncertainty on the background
  prediction.}
\end{figure}

\begin{figure*}[t]
\psfrag{m}{{\boldmath $M$}}
\begin{centering}
{\includegraphics[width=\columnwidth]{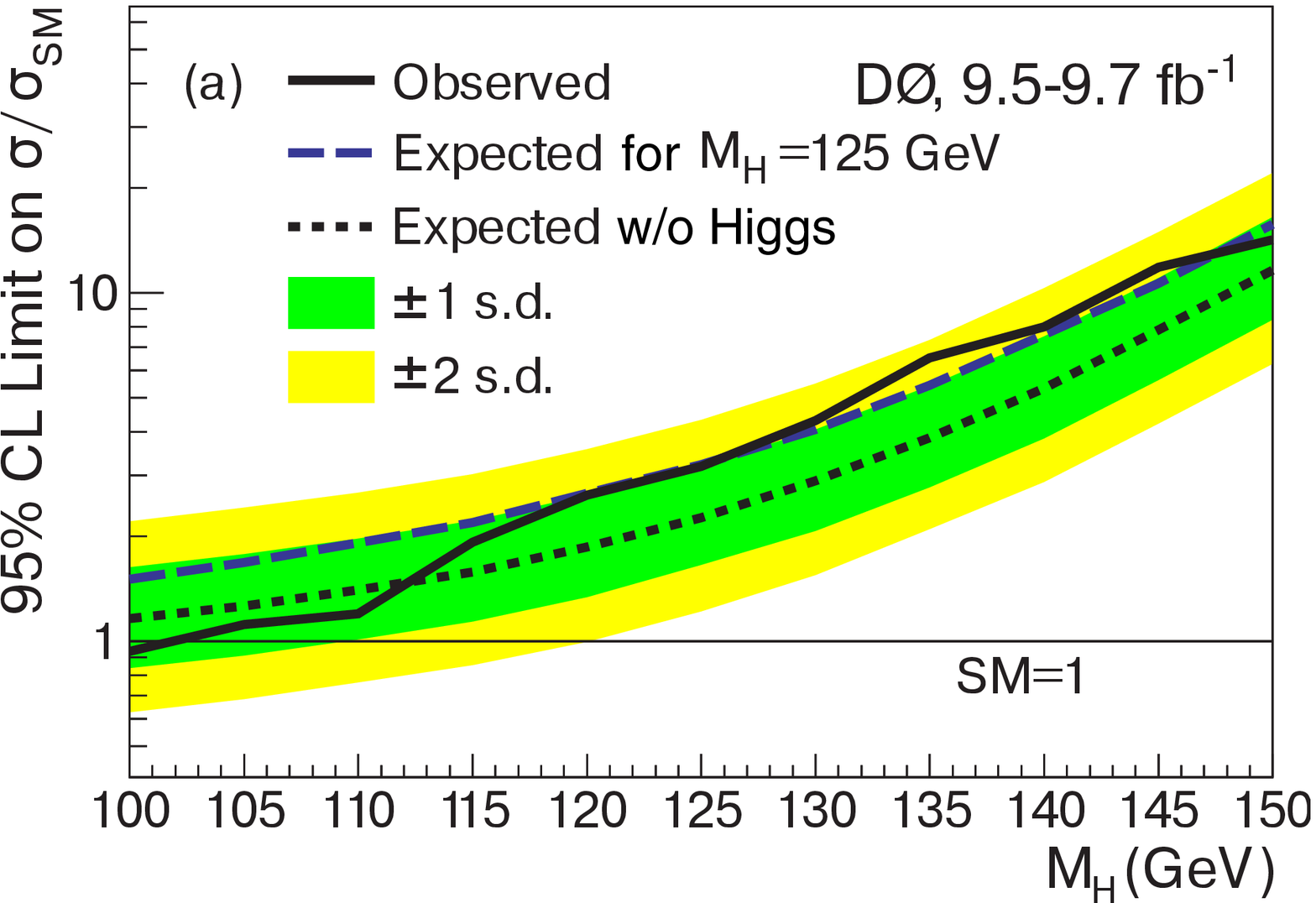}}
{\includegraphics[width=\columnwidth]{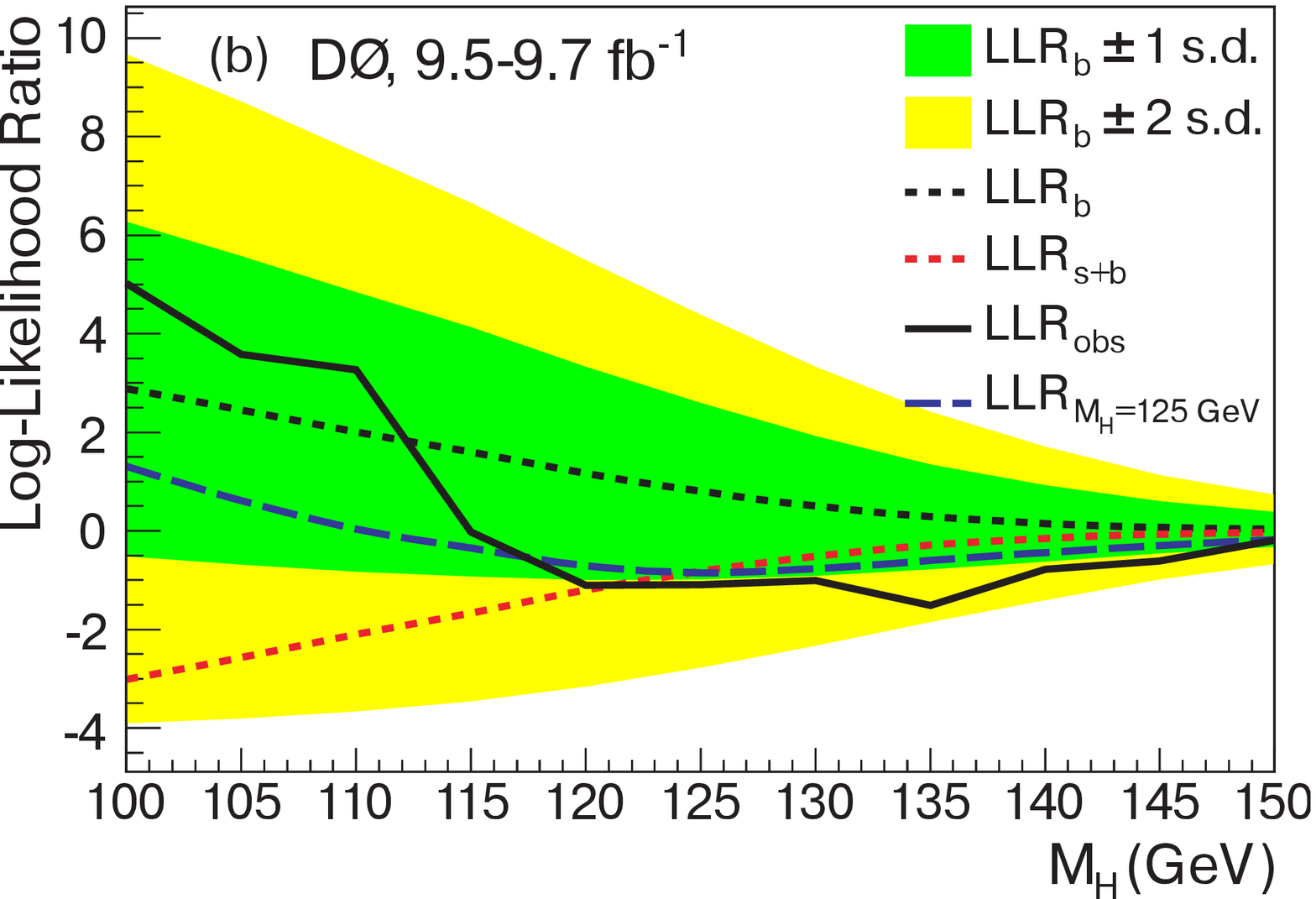}}
\caption{\label{fig:LLR} (color online).  (a) The 95\% C.L. cross
  section upper limit ratios versus $M_H$, and (b) LLR distribution
  versus $M_H$, for the combined $VH$, $H\to b\bar{b}$ analyses.  The
  solid lines represent the observed values in the data.  The
  short-dashed black (red) lines represent the median expected values
  under the background-only (signal-plus-background) hypothesis at
  each mass.  The long-dashed blue lines show the expected outcome
  from injecting a SM Higgs boson signal with $M_H=125$\gev.  The
  green and yellow shaded bands correspond to the regions enclosing 1
  and 2 s.d. variations about the median expected values under the
  background-only hypothesis, respectively.  }
\end{centering}
\end{figure*}

\begin{table*}[t]
\caption{Expected (median) and observed 95\% C.L. cross section upper
  limit ratios for the combined $VH$, $H\to b\bar{b}$ analyses over
  the $100\gev \leq M_H \leq 150\gev$ mass range.
\label{tab:limits}}
\begin{ruledtabular}
\begin{tabular}{lccccccccccc}
$M_{H}$ ($\rm Ge\kern -0.1em V$) &100 &105 &110 &115 &120 &125 &130 &135 &140 &145 &150 \\ \hline
Expected: &1.2 &1.3 &1.4 &1.6 &1.9 &2.3 &2.9 &3.8 &5.3 &7.8 &12 \\
Observed: &0.94 &1.1 &1.2 &1.9 &2.6 &3.2 &4.3 &6.5 &8.0 &12 &14 \\
\end{tabular}
\end{ruledtabular}
\end{table*}

To confirm the ability of these analyses to measure a signal and to
validate the background modeling, we perform a measurement of the $VZ$
production cross section in the same final states.  The only
difference from the Higgs boson search is to use SM $WZ$ and $ZZ$
production as the signal instead of $WH$ and $ZH$, while the rest of
the SM processes, including $WW$ production, are treated as
backgrounds. Multivariate discriminants using the same input variables
as in the Higgs boson searches are trained to separate the $VZ$ signal
from the backgrounds and the resulting distributions are fit to
determine the $VZ$ cross section.  The combination of all three
analyses yields $\sigma(VZ) = 3.3 \pm 1.4\thinspace$pb, consistent
with the SM prediction of $4.4 \pm 0.3$~pb~\cite{mcfm}.  The observed
(expected) significance of the measured excess is 2.5 (3.4) s.d.

The statistical analysis makes use of simultaneous fits to the
individual final discriminants, but it is useful for presentation
purposes to collect all of the inputs into a single distribution.
This is done by reordering the bins from the input distributions
according to their signal-to-background ratios ($s/b$), so that bins
with similar $\log_{10}(s/b)$ are combined.
Figure~\ref{fig:compare_subtracted} shows this distribution for the
$VZ$ cross section measurement and for the Higgs boson search with
$M_H = 125$\gev\ after subtracting the expected background from the
data.  The subtracted background corresponds to the maximum-likelihood
fit of the nuisance parameters to the data, and the posterior
uncertainty from that fit is also shown in the plot.

We derive limits on SM Higgs boson production $\sigma(VH) \times {\rm BR}(H
\rightarrow$$ b\bar{b})$ for Higgs boson masses in the range
$100\gev\leq M_H \leq 150\gev$ in steps of 5\gev.  We assume the
relative contributions of the different production and decay modes as
given by the SM prediction.  We present our results in terms of the
ratio of 95\% C.L. upper cross section limits to the SM predicted cross
section.  The SM prediction for Higgs boson production would therefore
be considered excluded at 95\% C.L. when this limit ratio falls below
unity.  Figure~\ref{fig:LLR}(a) shows the combined expected and
observed 95\% C.L. cross section limits as a ratio to the SM cross
section as a function of $M_H$. These results are also summarized in
Table~\ref{tab:limits}.  The LLR distributions for the combination are
shown in Fig.~\ref{fig:LLR}(b). Although consistent with the
background-only hypothesis for $M_H<115\gev$, the observed LLR
exhibits a signal-like excess at the level of 1--1.7 s.d. for the mass
range $120\gev\leq M_H \leq 145\gev$

To understand the compatibility of this excess with the hypothesis of
a SM Higgs boson, we obtain the best-fit cross section for the Higgs
boson signal relative to the SM prediction ($R^{\rm fit}$) as a
function of $M_H$.  This value is obtained by performing a
maximum-likelihood fit over all search channels simultaneously,
allowing the fit to vary all nuisance parameters within their priors
and with the Higgs boson cross section as a free parameter.
Figure~\ref{fig:xsec} shows the measured $\sigma(VH) \times {\rm
  BR}(H\to b\bar{b})$ as a function of $M_H$, including its $\pm 1$
s.d. uncertainty band, and compared with the SM prediction. At a mass
of $125\gev$, the best-fit cross section is $\sigma(VH) \times {\rm
  BR}(H\to b\bar{b}) = 140^{+140}_{-130}$~pb, which is
$1.2^{+1.2}_{-1.1}$ times the SM prediction.

\begin{figure}[h!]
\psfrag{m}{{\boldmath $M$}}
\begin{centering}
\includegraphics[width=\columnwidth]{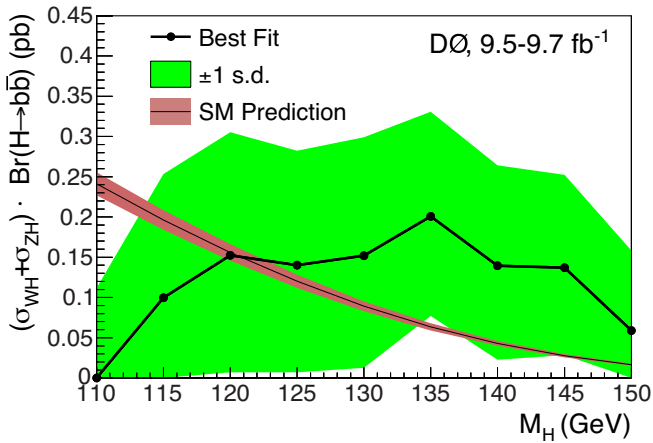}
\caption{\label{fig:xsec} (color online).  The best-fit value for
  $\sigma(VH) \times {\rm BR}(H\to b\bar{b})$ as a function of $M_H$.  The
  green shaded band corresponds to the 1 s.d. uncertainty around the
  best-fit cross section. Also shown is the SM prediction including
  the theoretical uncertainties.}
\end{centering}
\end{figure}

The significance of the data excess above the background prediction is
estimated by computing the $p$ value under the background-only
hypothesis using $R^{\rm fit}$ as the test statistic for each value of
$M_H$.  This $p$ value represents the probability to have a value of
$R^{\rm fit}$ as large or larger than that observed in the data due to a
background fluctuation.  The smallest $p$ value is obtained at a mass
of $135\gev$ and corresponds to a significance of $1.7$ s.d. above the
background-only prediction. This significance does not take into
account the look-elsewhere-effect~\cite{lee}, which accounts for
the possibility of a background fluctuation in the local $p$ value
anywhere in the tested mass range.  By taking into account existing
limits on $M_H$ in the $b\bar{b}$ decay mode~\cite{lepcombo}, the
search region becomes $115\gev\leq M_H \leq 150\gev$.  Given the
expected mass resolution of these searches of $\approx 16\%$, this
translates into a look-elsewhere-effect factor of $\approx 1.6$ for a global
significance of 1.5~s.d.  Also taking into account the existing SM
Higgs boson exclusions from the LHC~\cite{atlascomb2012,cmscomb2012}
experiments, there is no look-elsewhere-effect and we find an excess at $M_H=125\gev$
with a significance of 1.1~s.d.

In summary, we have presented a combination of searches for the SM
Higgs boson produced in association with a vector boson and decaying
into $b\bar{b}$, using the data sample collected with the D0 detector
in Run II of the Fermilab Tevatron Collider.  We achieve a sensitivity
that is competitive with other searches in this final
state~\cite{cdfbb,atlasbb,cmsbb}, deriving 95\% C.L. upper limits on the
Higgs boson cross section relative to the SM prediction in the mass
range $100\gev\leq M_H \leq 150\gev$ and excluding Higgs bosons with
masses smaller than $102\gev$ at the 95\% C.L.  In the mass range
$120\gev\leq M_H \leq 145\gev$, the data exhibit an excess above the
background prediction with a global significance of $1.5$ s.d. and a
magnitude consistent with that expected for the SM Higgs boson.

\input acknowledgement.tex
\clearpage 
\newpage

\end{document}
%

%% file: author_list.tex
\affiliation{LAFEX, Centro Brasileiro de Pesquisas F\'{i}sicas, Rio de Janeiro, Brazil}
\affiliation{Universidade do Estado do Rio de Janeiro, Rio de Janeiro, Brazil}
\affiliation{Universidade Federal do ABC, Santo Andr\'e, Brazil}
\affiliation{University of Science and Technology of China, Hefei, People's Republic of China}
\affiliation{Universidad de los Andes, Bogot\'a, Colombia}
\affiliation{Charles University, Faculty of Mathematics and Physics, Center for Particle Physics, Prague, Czech Republic}
\affiliation{Czech Technical University in Prague, Prague, Czech Republic}
\affiliation{Center for Particle Physics, Institute of Physics, Academy of Sciences of the Czech Republic, Prague, Czech Republic}
\affiliation{Universidad San Francisco de Quito, Quito, Ecuador}
\affiliation{LPC, Universit\'e Blaise Pascal, CNRS/IN2P3, Clermont, France}
\affiliation{LPSC, Universit\'e Joseph Fourier Grenoble 1, CNRS/IN2P3, Institut National Polytechnique de Grenoble, Grenoble, France}
\affiliation{CPPM, Aix-Marseille Universit\'e, CNRS/IN2P3, Marseille, France}
\affiliation{LAL, Universit\'e Paris-Sud, CNRS/IN2P3, Orsay, France}
\affiliation{LPNHE, Universit\'es Paris VI and VII, CNRS/IN2P3, Paris, France}
\affiliation{CEA, Irfu, SPP, Saclay, France}
\affiliation{IPHC, Universit\'e de Strasbourg, CNRS/IN2P3, Strasbourg, France}
\affiliation{IPNL, Universit\'e Lyon 1, CNRS/IN2P3, Villeurbanne, France and Universit\'e de Lyon, Lyon, France}
\affiliation{III. Physikalisches Institut A, RWTH Aachen University, Aachen, Germany}
\affiliation{Physikalisches Institut, Universit\"at Freiburg, Freiburg, Germany}
\affiliation{II. Physikalisches Institut, Georg-August-Universit\"at G\"ottingen, G\"ottingen, Germany}
\affiliation{Institut f\"ur Physik, Universit\"at Mainz, Mainz, Germany}
\affiliation{Ludwig-Maximilians-Universit\"at M\"unchen, M\"unchen, Germany}
\affiliation{Fachbereich Physik, Bergische Universit\"at Wuppertal, Wuppertal, Germany}
\affiliation{Panjab University, Chandigarh, India}
\affiliation{Delhi University, Delhi, India}
\affiliation{Tata Institute of Fundamental Research, Mumbai, India}
\affiliation{University College Dublin, Dublin, Ireland}
\affiliation{Korea Detector Laboratory, Korea University, Seoul, Korea}
\affiliation{CINVESTAV, Mexico City, Mexico}
\affiliation{Nikhef, Science Park, Amsterdam, Netherlands}
\affiliation{Radboud University Nijmegen, Nijmegen, Netherlands}
\affiliation{Joint Institute for Nuclear Research, Dubna, Russia}
\affiliation{Institute for Theoretical and Experimental Physics, Moscow, Russia}
\affiliation{Moscow State University, Moscow, Russia}
\affiliation{Institute for High Energy Physics, Protvino, Russia}
\affiliation{Petersburg Nuclear Physics Institute, St. Petersburg, Russia}
\affiliation{Instituci\'{o} Catalana de Recerca i Estudis Avan\c{c}ats (ICREA) and Institut de F\'{i}sica d'Altes Energies (IFAE), Barcelona, Spain}
\affiliation{Uppsala University, Uppsala, Sweden}
\affiliation{Lancaster University, Lancaster LA1 4YB, United Kingdom}
\affiliation{Imperial College London, London SW7 2AZ, United Kingdom}
\affiliation{The University of Manchester, Manchester M13 9PL, United Kingdom}
\affiliation{University of Arizona, Tucson, Arizona 85721, USA}
\affiliation{University of California Riverside, Riverside, California 92521, USA}
\affiliation{Florida State University, Tallahassee, Florida 32306, USA}
\affiliation{Fermi National Accelerator Laboratory, Batavia, Illinois 60510, USA}
\affiliation{University of Illinois at Chicago, Chicago, Illinois 60607, USA}
\affiliation{Northern Illinois University, DeKalb, Illinois 60115, USA}
\affiliation{Northwestern University, Evanston, Illinois 60208, USA}
\affiliation{Indiana University, Bloomington, Indiana 47405, USA}
\affiliation{Purdue University Calumet, Hammond, Indiana 46323, USA}
\affiliation{University of Notre Dame, Notre Dame, Indiana 46556, USA}
\affiliation{Iowa State University, Ames, Iowa 50011, USA}
\affiliation{University of Kansas, Lawrence, Kansas 66045, USA}
\affiliation{Kansas State University, Manhattan, Kansas 66506, USA}
\affiliation{Louisiana Tech University, Ruston, Louisiana 71272, USA}
\affiliation{Boston University, Boston, Massachusetts 02215, USA}
\affiliation{Northeastern University, Boston, Massachusetts 02115, USA}
\affiliation{University of Michigan, Ann Arbor, Michigan 48109, USA}
\affiliation{Michigan State University, East Lansing, Michigan 48824, USA}
\affiliation{University of Mississippi, University, Mississippi 38677, USA}
\affiliation{University of Nebraska, Lincoln, Nebraska 68588, USA}
\affiliation{Rutgers University, Piscataway, New Jersey 08855, USA}
\affiliation{Princeton University, Princeton, New Jersey 08544, USA}
\affiliation{State University of New York, Buffalo, New York 14260, USA}
\affiliation{University of Rochester, Rochester, New York 14627, USA}
\affiliation{State University of New York, Stony Brook, New York 11794, USA}
\affiliation{Brookhaven National Laboratory, Upton, New York 11973, USA}
\affiliation{Langston University, Langston, Oklahoma 73050, USA}
\affiliation{University of Oklahoma, Norman, Oklahoma 73019, USA}
\affiliation{Oklahoma State University, Stillwater, Oklahoma 74078, USA}
\affiliation{Brown University, Providence, Rhode Island 02912, USA}
\affiliation{University of Texas, Arlington, Texas 76019, USA}
\affiliation{Southern Methodist University, Dallas, Texas 75275, USA}
\affiliation{Rice University, Houston, Texas 77005, USA}
\affiliation{University of Virginia, Charlottesville, Virginia 22904, USA}
\affiliation{University of Washington, Seattle, Washington 98195, USA}
\author{V.M.~Abazov} \affiliation{Joint Institute for Nuclear Research, Dubna, Russia}
\author{B.~Abbott} \affiliation{University of Oklahoma, Norman, Oklahoma 73019, USA}
\author{B.S.~Acharya} \affiliation{Tata Institute of Fundamental Research, Mumbai, India}
\author{M.~Adams} \affiliation{University of Illinois at Chicago, Chicago, Illinois 60607, USA}
\author{T.~Adams} \affiliation{Florida State University, Tallahassee, Florida 32306, USA}
\author{G.D.~Alexeev} \affiliation{Joint Institute for Nuclear Research, Dubna, Russia}
\author{G.~Alkhazov} \affiliation{Petersburg Nuclear Physics Institute, St. Petersburg, Russia}
\author{A.~Alton$^{a}$} \affiliation{University of Michigan, Ann Arbor, Michigan 48109, USA}
\author{G.~Alverson} \affiliation{Northeastern University, Boston, Massachusetts 02115, USA}
\author{A.~Askew} \affiliation{Florida State University, Tallahassee, Florida 32306, USA}
% ...
%
%
\collaboration{D0 Collaboration\footnote{with visitors from
$^{a}$Augustana College, Sioux Falls, SD, USA,
$^{b}$The University of Liverpool, Liverpool, UK,
$^{c}$UPIITA-IPN, Mexico City, Mexico,
$^{d}$DESY, Hamburg, Germany,
$^{e}$SLAC, Menlo Park, CA, USA,
$^{f}$University College London, London, UK,
$^{g}$Centro de Investigacion en Computacion - IPN, Mexico City, Mexico,
$^{h}$ECFM, Universidad Autonoma de Sinaloa, Culiac\'an, Mexico
and
$^{i}$Universidade Estadual Paulista, S\~ao Paulo, Brazil.
}} \noaffiliation
\vskip 0.25cm

%% file: acknowledgement.tex
%
We thank the staffs at Fermilab and collaborating institutions,
and acknowledge support from the
DOE and NSF (USA);
CEA and CNRS/IN2P3 (France);
MON, NRC KI and RFBR (Russia);
CNPq, FAPERJ, FAPESP and FUNDUNESP (Brazil);
DAE and DST (India);
Colciencias (Colombia);
CONACyT (Mexico);
NRF (Korea);
FOM (Netherlands);
STFC and the Royal Society (United Kingdom);
MSMT and GACR (Czech Republic);
BMBF and DFG (Germany);
SFI (Ireland);
The Swedish Research Council (Sweden);
and
CAS and CNSF (China).